# Thermal conductivity and Lorenz ratio of metals at intermediate temperature: a first-principles analysis

Shouhang Li[1], Zhen Tong[1,2], Xinyu Zhang[1] and Hua Bao[1,*]

[1]University of Michigan-Shanghai Jiao Tong University Joint Institute, Shanghai Jiao Tong University, Shanghai 200240, P. R. China

[2] Shenzhen JL Computational Science and Applied Research Institute, Shenzhen 518109, P. R. China

**abstract**

Electronic and phononic thermal conductivity are involved in the thermal conduction for metals and Wiedemann-Franz law is usually employed to predict them separately. However, Wiedemann-Franz law is shown to be invalid at intermediate temperatures. Here, to obtain the accurate thermal conductivity and Lorenz ratio for metals, the momentum relaxation time is used for electrical conductivity and energy relaxation time for electronic thermal conductivity. The mode-level first-principles calculation is conducted on two representative metals copper and aluminum. It is shown that the method can correctly predict electrical transport coefficients from 6 to 300 K. Also, the anomalous Lorenz ratio is observed within the present scheme, which has significant departure from the Sommerfeld value. The calculation scheme can be expanded to other metallic systems and is valuable in a better understanding of the electron dynamics and transport properties of metals.

[*] To whom correspondence should be addressed. Email: hua.bao@sjtu.edu.cn (HB)

# I. Introduction

Heat conduction in metals involves complicated electron and phonon transport and scattering processes, which has historically been a research focus in solid-state physics [1,2]. The Wiedemann-Franz law states that the ratio of electronic thermal conductivity ($\kappa_{el}$) to electrical conductivity ($\sigma$) is proportional to the absolute temperature $T$ [3], and plays a vital role in the understanding of thermal transport in metals. The proportionality constant (also known as the Lorenz ratio) is often taken as Sommerfeld value, $L_0 = 2.44 \times 10^{-8} \, \mathrm{W\Omega/K^2}$. It has been well recognized that the Lorenz ratio is generally similar to Sommerfeld value at low or high temperature range [4], in which the elastic phonon-impurity scattering or nearly elastic electron-phonon scattering prevails. However, at the intermediate temperature range ($0.1\,\Theta_D \sim \Theta_D$, with $\Theta_D$ being the Debye temperature) where inelastic electron-phonon scattering dominates, the Lorenz ratio can significantly deviate from the Sommerfeld value and the mechanism is worth further exploring [5].

Although experimental measurements of electrical and thermal conductivity are available for typical elemental metals at the intermediate temperature range [6], to correctly obtain electronic components of thermal conductivity is still challenging. In most experiments, the phonon thermal conductivity is simply neglected when calculating the Lorenz ratio [7]. In order to explicitly obtain electron and phonon contributions to thermal conductivity, one either needs to apply large magnetic field to suppress electron transport [8,9], or needs the complicated alloying method to extrapolate a series of samples with different concentrations [10]. Due to these difficulties, the experimental data is only available for a handful of simple metals [8,10,11]. On the other hand, to understand the underlining physics, the Bloch-Grüneisen (BG) model [5,12,13] developed in 1930s is still widely applied to explain the deviation of Lorenz ratio for the intermediate temperature range [8,14]. The decreasing trend with temperature for transport coefficients and Lorenz ratio can be partially captured by the BG model [5]. Nevertheless, the BG model is based on the assumptions of free electrons, Debye phonon spectrum, and that electrons only scatter with longitudinal acoustic phonons. Also, the inelastic electron-phonon scattering is treated with the ideal spherical Fermi surface assumption [5]. Therefore, it is difficult to obtain thermal conductivity and Lorenz ratio for metals at intermediate temperatures in quantitative manner.

Recent advances in first-principles calculations allow to quantitatively determine the transport coefficients of metals. Early calculations adopt constant electron relaxation time approximation combing first-principles electron band structure [15], which can introduce large deviations even in simple elemental metals [16,17]. Moreover, due to the assumption of constant relaxation time, Wiedemann-Franz law must be valid in the whole temperature range in free electron metals [5]. Allen's model, which is the lowest order variational approximation of the solution for electron Boltzmann transport equation (BTE) [18-20], allows to obtain the transport coefficients in real metallic systems. Nevertheless, non-negligible deviations still exist in some elemental metals in Allen's model [21,22]. Recently, the accurate mode-level calculation of electron-phonon scattering is enabled by employing Maximally Localized Wannier Function (MLWF) interpolation technique [23]. Using this approach, explicit electron-phonon scattering rate can be extracted and substituted into the electron BTE. Combining with anharmonic lattice dynamics, the phonon and electron thermal conductivities can be separately calculated. This method has been successfully applied to element metals [24-26], intermetallics [27], doped semiconductors [28-30], and intrinsic semiconductor at high temperature [31]. All these calculations are conducted at temperatures similar to or higher than the Debye temperature. As a result, the predicted Lorenz ratios in metals are quite consistent with the Sommerfeld value [25,26]. The mechanism of inelastic electron-phonon scattering and its effect on electron transport in metals at intermediate temperatures have never been carefully explored.

In this work, the thermal transport properties and Lorenz ratio at intermediate temperatures (6-300 K) for metals are studied from first principles with a detailed analysis of inelastic electron-phonon scattering. Two representative metals copper (Cu) and aluminum (Al) are considered. We first review the transport theory within electron BTE and suggest that different relaxation times are necessary to describe the electrical transport and thermal transport. Then, convergence issues for the transport coefficients with respect to smearing parameter and mesh size in Brillouin zone integration are carefully examined. We point out the importance of these parameters to the correct prediction of the transport coefficients. Then, the electrical resistivity conductivity and electronic thermal conductivity are presented and compared with existing models. The mechanism of inelastic scattering and its effect on transport coefficients and Lorenz ratio at intermediate temperature range are also discussed.

## II. Theory and Methods

The calculation of electron transport properties can be conducted with the framework of electron BTE. With both the external electrical field and temperature gradient, the steady-state linearized electron BTE can be deduced as

$$\left(-\frac{\partial f_{n\mathbf{k}}^0}{\partial \varepsilon_{n\mathbf{k}}}\right)\vec{v}_{n\mathbf{k}} \cdot \left[e\vec{\mathcal{E}} + \frac{\varepsilon_{n\mathbf{k}} - \varepsilon_F}{T}\nabla T\right] = \left(\frac{\partial f_{n\mathbf{k}}}{\partial t}\right)_{coll} \tag{1}$$

where $f_{n\mathbf{k}}^0$ is electron equilibrium distribution at the electron mode $n\mathbf{k}$ ($n$-band index, $\mathbf{k}$-wave vector), namely Fermi-Dirac distribution. $e$ is the elementary charge, $\vec{\mathcal{E}}$ is the electrical field, $\varepsilon$ is the electron energy, $\varepsilon_F$ is Fermi energy, and $\vec{v}$ is the electron group velocity. The two terms on the left-hand side express the deviations from the equilibrium generated by electrical field and temperature gradient, respectively. The right-hand side is the electron collision term, which returns the system to equilibrium. Under the relaxation time approximation, the collision term is usually simplified as Eq. (2), where $\tau_{n\mathbf{k}}$ is the relaxation times and it quantifies how quickly the electron returns to equilibrium [5].

$$\left(\frac{\partial f_{n\mathbf{k}}}{\partial t}\right)_{coll} = -\frac{f_{n\mathbf{k}} - f_{n\mathbf{k}}^0}{\tau_{n\mathbf{k}}} = -\frac{\delta f_{n\mathbf{k}}}{\tau_{n\mathbf{k}}} \tag{2}$$

In most works [24,27,29], $\tau_{n\mathbf{k}}$ is regarded to be the same for both electrical transport and thermal transport process. However, recent work [4] proposed that the relaxation times for charge transport and energy transport should be distinct from each other in order to accurately capture the transport properties and Lorenz ratio at intermediate temperature. Therefore, the deviation of electron distribution with relaxation time approximation can be expressed as [4,5]

$$\delta f_{n\mathbf{k}} = \tau_{\sigma,n\mathbf{k}}(T) e \vec{v}_{n\mathbf{k}} \cdot \vec{\mathcal{E}} \left( \frac{\partial f_{n\mathbf{k}}^0}{\partial \varepsilon_{n\mathbf{k}}} \right) + \tau_{\kappa,n\mathbf{k}}(T) \frac{\varepsilon(\mathbf{k}) - \varepsilon_F}{T} \vec{v}_{n\mathbf{k}} \cdot \nabla T \left( \frac{\partial f_{n\mathbf{k}}^0}{\partial \varepsilon_{n\mathbf{k}}} \right), \quad (3)$$

in which the $\tau_{\sigma,n\mathbf{k}}$ is the electrical relaxation time (also called momentum relaxation time) and $\tau_{\kappa,n\mathbf{k}}$ is thermal relaxation time (also called energy relaxation time). If we only consider the electron-phonon scattering process, the energy relaxation time can be determined from an integration of all scattering process for the electron at $\varepsilon_{n\mathbf{k}}$ and expressed as

$$\frac{1}{\tau_{n\mathbf{k}}} = \frac{2\pi}{\hbar} \sum_{m\mathbf{k}+\mathbf{q}} |g_{mn,\nu}(\mathbf{k},\mathbf{q})|^2 \left\{ \begin{array}{l} \left[ n_{\mathbf{q}\nu}^0(T) + f_{m\mathbf{k}+\mathbf{q}}^0(T) \right] \delta \left( \varepsilon_{n\mathbf{k}} - \varepsilon_{m\mathbf{k}+\mathbf{q}} + \hbar\omega_{\mathbf{q}\nu} \right) \\ + \left[ n_{\mathbf{q}\nu}^0(T) + 1 - f_{m\mathbf{k}+\mathbf{q}}^0(T) \right] \delta \left( \varepsilon_{n\mathbf{k}} - \varepsilon_{m\mathbf{k}+\mathbf{q}} - \hbar\omega_{\mathbf{q}\nu} \right) \end{array} \right\} \quad (4)$$

where $n_{\mathbf{q}\nu}^0$ is equilibrium Bose-Einstein distribution related to phonon $\mathbf{q}\nu$. The first and second term in the curly brace is related to the absorption and emission process during electron-phonon coupling, respectively. The electron-phonon coupling matrix element [32] can be expressed as

$$g_{mn,\nu}(\mathbf{k},\mathbf{q}) = \left( 2\omega_{\mathbf{q}\nu} \right)^{1/2} \left\langle \psi_{m\mathbf{k}+\mathbf{q}} | \Delta V_{\mathbf{q}\nu} | \psi_{n\mathbf{k}} \right\rangle \quad (5)$$

where $\omega$ is phonon frequency. $\psi_{n\mathbf{k}}$ and $\psi_{m\mathbf{k}+\mathbf{q}}$ are the initial and final Bloch electron states of the scattering process, respectively.

The momentum relaxation time at $\varepsilon_{n\mathbf{k}}$ is given by

$$\frac{1}{\tau_{\sigma,n\mathbf{k}}} = \sum_{m\mathbf{k}+\mathbf{q}} \frac{1}{\tau_{\kappa,n\mathbf{k},m\mathbf{k}+\mathbf{q}}} \left(1 - \cos\theta_{n\mathbf{k},m\mathbf{k}+\mathbf{q}}\right) \tag{6}$$

The two relaxation times in a scattering process differ by an efficiency factor $\alpha_{n\mathbf{k},m\mathbf{k}+\mathbf{q}} = \left(1 - \cos\theta_{n\mathbf{k},m\mathbf{k}+\mathbf{q}}\right)$, which goes in the range of 0 to 2. Here $\theta_{n\mathbf{k},m\mathbf{k}+\mathbf{q}}$ is the scattering angle between the electron states $n\mathbf{k}$ and $m\mathbf{k}+\mathbf{q}$. With the assumption of $\tau_{n\mathbf{k}}\vec{v}_{n\mathbf{k}} \approx \tau_{m\mathbf{k}+\mathbf{q}}\vec{v}_{m\mathbf{k}+\mathbf{q}}$ [33], we have

$$\cos\theta_{n\mathbf{k},m\mathbf{k}+\mathbf{q}} = \frac{\vec{v}_{m\mathbf{k}+q} \cdot \vec{v}_{n\mathbf{k}}}{\left|\vec{v}_{n\mathbf{k}}\right|\left|\vec{v}_{m\mathbf{k}+q}\right|}. \tag{7}$$

The efficiency factor arises from the fact that the scattering of electrical current is efficient only when the direction of electron motion is changed. While the energy exchange happens in any scattering process and thus the scattering efficiency for heat current is always effective. Note that in many previous works [24,25,29], the difference between momentum relaxation time and energy relaxation time is not considered and the efficiency factor is neglected in both calculations. This may not cause much error at high-temperature range since the large-angle scatterings dominate the scattering process. However, as will be shown later, at lower temperatures, the correct relaxation times must be employed in order to obtain accurate transport properties and Lorenz ratios.

As electrical current ($\vec{J}$) and heat current ($\vec{Q}$) can be expressed as [5]

$$\begin{cases} \vec{J} = \sum_{n\mathbf{k}} ev_{n\mathbf{k}} f_{n\mathbf{k}} = \sum_{n\mathbf{k}} ev_{n\mathbf{k}} \left(f_{n\mathbf{k}}^0 + \delta f_{n\mathbf{k}}\right) \\ \vec{Q} = \sum_{n\mathbf{k}} \left(\varepsilon_{n\mathbf{k}} - \varepsilon_F\right) v_{n\mathbf{k}} f_{n\mathbf{k}} = \sum_{n\mathbf{k}} \left(\varepsilon_{n\mathbf{k}} - \varepsilon_F\right) v_{n\mathbf{k}} \left(f_{n\mathbf{k}}^0 + \delta f_{n\mathbf{k}}\right) \end{cases} \tag{8}$$

Eq. (3) can be entered into Eq. (8) and we have,

$$\begin{cases} \vec{J} = L_{EE}\vec{\varepsilon} + L_{ET}\nabla T \\ \vec{Q} = L_{TE}\vec{\varepsilon} + L_{TT}\nabla T \end{cases}, \tag{9}$$

with the coefficients given by

$$\begin{cases} L_{EE} = -\dfrac{e^2}{N_{\mathbf{k}}V}\sum_{m\mathbf{k}} v_{m\mathbf{k},\alpha} v_{m\mathbf{k},\beta} \tau_{\sigma,m\mathbf{k}}(\varepsilon_F,T) \dfrac{\partial f^0(\varepsilon_{m\mathbf{k}},\varepsilon_F,T)}{\partial \varepsilon_{m\mathbf{k}}} \\ L_{ET} = -\dfrac{e}{N_{\mathbf{k}}VT}\sum_{m\mathbf{k}} v_{m\mathbf{k},\alpha} v_{m\mathbf{k},\beta} \tau_{\kappa,m\mathbf{k}}(\varepsilon_F,T)(\varepsilon-\varepsilon_F) \dfrac{\partial f^0(\varepsilon_{m\mathbf{k}},\varepsilon_F,T)}{\partial \varepsilon_{m\mathbf{k}}} \\ L_{TE} = -\dfrac{e}{N_{\mathbf{k}}V}\sum_{m\mathbf{k}} v_{m\mathbf{k},\alpha} v_{m\mathbf{k},\beta} \tau_{\sigma,m\mathbf{k}}(\varepsilon_F,T)(\varepsilon-\varepsilon_F) \dfrac{\partial f^0(\varepsilon_{m\mathbf{k}},\varepsilon_F,T)}{\partial \varepsilon_{m\mathbf{k}}} \\ L_{TT} = \dfrac{1}{N_{\mathbf{k}}V}\sum_{m\mathbf{k}} -\dfrac{(\varepsilon_{m\mathbf{k}}-\varepsilon_F)^2}{T} v_{m\mathbf{k},\alpha} v_{m\mathbf{k},\beta} \tau_{\kappa,m\mathbf{k}}(\varepsilon_F,T) \dfrac{\partial f_{FD}(\varepsilon_{m\mathbf{k}},\varepsilon_F,T)}{\partial \varepsilon_{m\mathbf{k}}} \end{cases}. \tag{10}$$

$\alpha$ and $\beta$ are Cartesian coordinate components. $N_{\mathbf{k}}$ is the total number of **k**-points in the first Brillouin zone. $V$ is the volume of the unit cell.

From Eq. (9), the electrical and thermal conductivity can be expressed from the definitions of the two transport coefficients,

$$\begin{cases} \sigma = L_{EE} \\ \kappa_{\mathrm{el}} = -\left(L_{TT} - \dfrac{L_{TE}L_{ET}}{L_{EE}}\right) \end{cases} \tag{11}$$

The term $\dfrac{L_{TE}L_{ET}}{L_{EE}}$ in the expression of electronic thermal conductivity is very small in metals and can be directly ignored. Hence only energy relaxation time $\tau_\kappa$ is contained in electronic thermal conductivity in our later calculation. Note that for $L_{EE}$, the momentum relaxation time is adopted, while for $L_{TT}$, the energy relaxation time is used. This is different from many references [24,25,29,30].

In the subsequent discussions, we consider two common metals, copper (Cu) and aluminum (Al) as examples. The first-principles calculations are carried out with Quantum Espresso [34]. A Perdew-Burke-Ernzerhof (PBE) form [35] of generalized gradient approximation (GGA) is employed as the exchange-correlation functional. The cutoff energy of the plane wave is set as 180 Ry for Cu and 100 Ry for Al to ensure convergence, and the convergence threshold of electron energy is set to be $10^{-10}$ Ry for the self-consistent field calculation.

The lattice vectors and atomic positions are fully relaxed based on the Broyden-Fretcher-Goldfarb-Shanno optimization method [36-39]. The optimized lattice constants come out to be 3.669 Å and 4.043 Å (experimental values are 3.6147 Å and 4.0496 Å) [40] for Cu and Al with face-centered cubic (FCC) lattice, respectively. For the electronic properties calculations, the two categories of electron-phonon scattering rates are calculated by our in-house modified Electron-Phonon Wannier (EPW) package [32]. The phonon spectrum and phonon potential differential are obtained on $6\times6\times6$ **q**-points mesh and the spectrums match experimental well, as shown in Appendix A. The electron-phonon coupling matrix elements are first calculated on the coarse grids of $12\times12\times12$ **k**-points and $6\times6\times6$ **q**-points, and are then interpolated to the sufficiently dense **k**-points and **q**-points to ensure the convergence of electrical transport coefficients in the whole temperature range. The electron band structures calculated under the Wannier scheme match Density Function Theory (DFT) band structures quite well, shown in Appendix A.

To calculate the phonon thermal conductivity component, we employed the widely used anharmonic lattice dynamics scheme to calculate the phonon-phonon scattering process [41]. The harmonic force constant is obtained employing the density-functional perturbation theory [42] under $6\times6\times6$ **q**-points mesh. The cubic force constant is extracted with THIRDORDER.PY package [41]. A supercell of $4\times4\times4$ was used and the 3$^{rd}$ nearest neighbors were included for the third-order interactions. A $3\times3\times3$ **k**-points mesh was used and the convergence is ensured. The iterative solution of phonon BTE is adopted. Note that in metals, phonon-electron scattering must also be considered. The calculation method of phonon-electron scattering and phonon thermal conductivity has been described in our previous works [25] and thus not repeated here.

# III. Results and Discussions

## A. Convergence study

In order to obtain correct relaxation times in Eqs. (4) and (6) numerically, Brillouin zone integration needs to be performed. Unlike the calculations at high temperature, the scattering process at intermediate temperature is weak and very dense Brillouin zone sampling must be performed in order to obtain the converged electron relaxation times [43]. The approximation form employed for the $\delta$ function in Eq. (4) is also a curial issue in the convergence of electron scattering rates. In this work, we follow Ref. [32] and adopt the Gaussian broadening scheme as $\delta(x) = \lim_{\eta \to 0} \frac{1}{\sqrt{\pi}} \frac{1}{\eta} e^{-\left(\frac{x}{\eta}\right)^2}$. The choice of broadening parameter $\eta$ must be balanced with the density of **q**- and **k**-points sampling. Usually, a small enough $\eta$ compatible with **q**- and **k**-points mesh is used. As will be shown below, inappropriate usage of the Gaussian broadening parameter can give unphysical results.

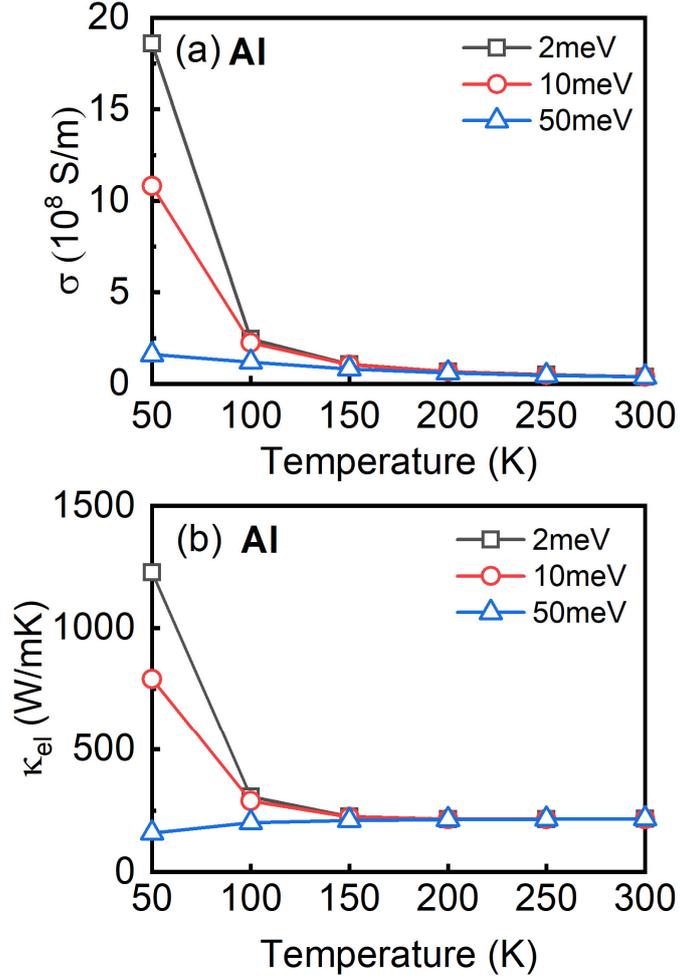

FIG. 1. Variations of (a) electrical conductivity and (b) electronic thermal conductivity of Al in the temperature range of 50 to 300 K with three Gaussian broadening smearing parameters 2, 10 and 50 meV. The **q**- and **k**-points mesh are set as $60 \times 60 \times 60$ and $80 \times 80 \times 80$, respectively.

We predict the electrical and thermal conductivity with three different $\eta$ values, 2, 10 and 50 meV in Al. The **q**- and **k**-points mesh are set as relatively dense values, $60 \times 60 \times 60$ and $80 \times 80 \times 80$, respectively. As it is shown in FIG. 1, there is no much difference in both electrical and thermal conductivity with different $\eta$ when the temperature is larger than 200 K. However, significant deviations appear in both electrical and thermal conductivity as the temperature is below 150 K. When the broadening parameter is set to be 50 meV, the electronic thermal conductivity displays and increasing trend with temperature, which is in contradict with experimental observations [6]. Note such unphysical phenomenon is observed in previous calculations [24]. Therefore, to accurately obtain the converged electron scattering rates, smaller $\eta$ should be used as the smearing of Fermi-Dirac distribution will be smaller with temperature decreasing. In this work, the value of $k_B T$ for $\eta$ is used ($k_B$ is the Boltzmann constant), which can yield correct results.

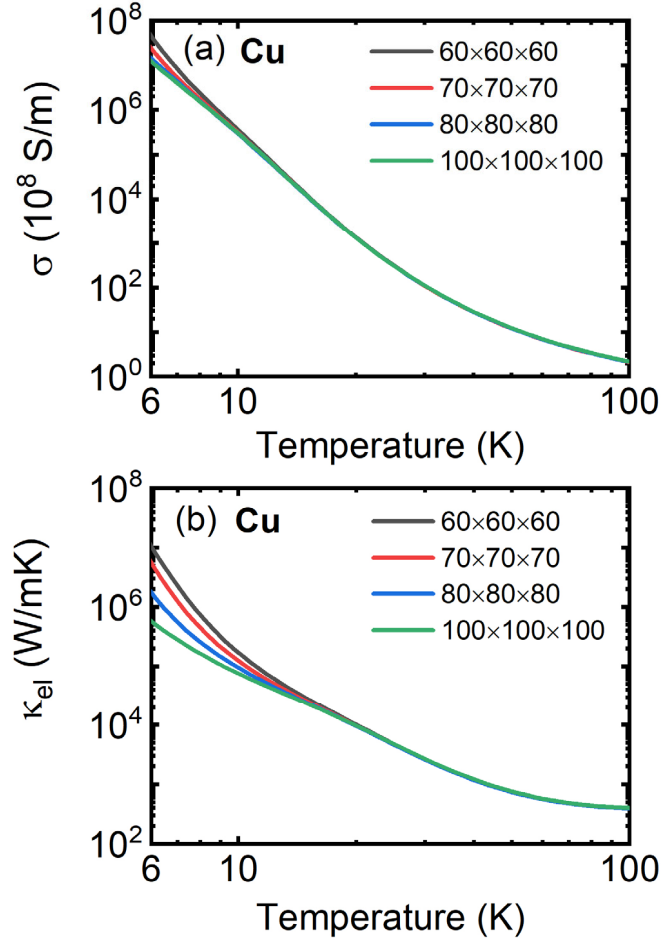

FIG. 2. Calculated (a) electrical conductivity and (b) electronic thermal conductivity of Cu with respect to the **q**-points mesh size in the temperature range of 6 to 100 K. The **k**-mesh size is kept as $200 \times 200 \times 200$ in all the calculations.

Because small Gaussian broadening parameters are employed at low temperatures, an ultra-dense **q**-points mesh should be used to ensure the converged electron scattering rates. Here we present the convergence test with different **q**-points mesh for both electrical conductivity and electronic thermal conductivity of Cu, as shown in FIG. 2. An ultra-dense **k**-points mesh of $200 \times 200 \times 200$ is fixed for all cases, as it is dense enough to capture the variation of scattering rates near the Fermi surface. Both electrical conductivity and electronic thermal conductivity get converged easier as the temperature is larger than 20 K, where we find $60 \times 60 \times 60$ **q**-points mesh is enough to achieve convergence. In comparison, for temperature below 20 K, an extremely dense **q**-point mesh of $100 \times 100 \times 100$ is sufficiently large to ensure the convergence for electrical conductivity but is not enough for thermal conductivity. However, considering even denser **q**-points mesh becomes computationally formidable. Therefore, we use $100 \times 100 \times 100$ **q**-points mesh and $200 \times 200 \times 200$ **k**-points mesh in the subsequent calculations. Note that small **q**-points mesh would result in an underestimation for the two transport coefficients due to the insufficient scattering phase space accessible for electron modes.

The converged electron scattering rates corresponding to momentum ($1/\tau_\sigma$) and energy ($1/\tau_\kappa$) relaxation times with different temperatures for Cu and Al are shown in FIG. 3. The scattering rates increase significantly with temperature, which is mainly due to the increase of phonon distribution $n^0$. A drastic variation appears near the Fermi surface. There is a sharp decrease of around five orders magnitude for $1/\tau_\sigma$ and around four orders magnitude for $1/\tau_\kappa$ as the temperature goes from 100 K to 10 K for Cu. It clearly demonstrates that the constant electron relaxation time assumption [15,44] may give large uncertainty for calculating the electrical transport properties and the mode-level calculation is necessary. The difference between the scattering rates is small at 50 K and 100 K for both Cu and Al. But the difference becomes large at 10 K, which is about one order of magnitude in Cu. There are valleys near the Fermi energy as the temperature is smaller than 100 K. It originates from the decreasing scattering phase space with the suppression of long wavevector phonon modes as the temperature decreases. A similar phenomenon is also observed in the calculation for Pb [43].

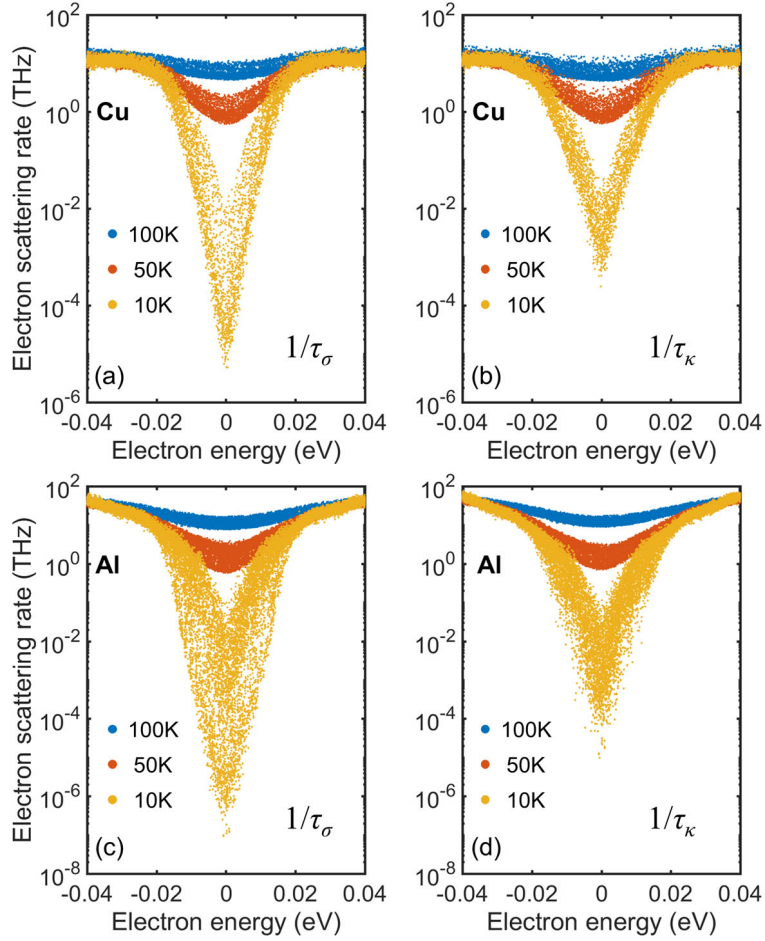

FIG. 3. The electron scattering rates corresponding to momentum relaxation time with different temperatures for (a) Cu and (c) Al. The electron scattering rates corresponding to energy relaxation time with different temperatures for (b) Cu and (d) Al. The electron energy is normalized to Fermi energy ($\varepsilon_F$).

## B. Electrical transport coefficients

The electrical conductivities for Cu and Al calculated by MRTA, ERTA, and Allen's model are compared to experimental data and shown in FIG. 4. The electrical conductivities are found to decrease with the increase of temperature. This is related to the increased electron-phonon scattering with temperature. Overall, the electrical conductivity predicted by momentum relaxation time approximation (MRTA) is closer to the experimental data [45] in the whole temperature range. Since the energy relaxation time approximation (ERTA) is also employed in electrical conductivity prediction in previous works [25,46,47], we also calculate the electrical conductivity using the energy relaxation time. It is found that the electrical conductivity predicted by ERTA is very close to that evaluated by MRTA as the temperature is larger than 40 K for Cu. However, the difference becomes significant as the temperature is below 40 K. This arises from the difference between the two relaxation times, as later shown in FIG. 6(a). The momentum relaxation time is larger than energy relaxation time and finally results in the smaller electrical conductivity in ERTA. It should be noted that the difference between ERTA and MRTA is noticeable in Al even in the temperature of 100-300K, which is also observed in the previous study [48]. It demonstrates that ERTA is inappropriate in the prediction of electrical conductivity for Al even at room temperature.

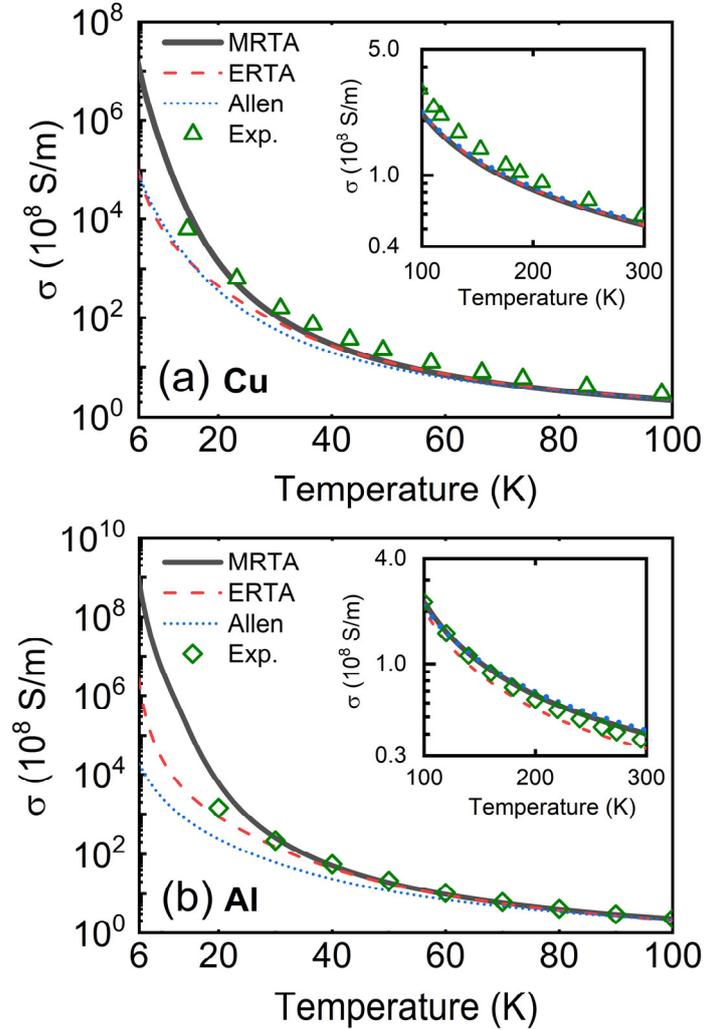

FIG. 4. The electrical conductivity for Cu (a) and Al (c) calculated with MRTA (black solid lines), ERTA (red dash lines) and Allen's model (blue dot lines). The experimental data are taken from Ref. [45]. The insets of (a) and (b) are the electrical conductivity in the temperature range of 100-300 K. The ideal experimental data is obtained by subtracting the residual resistance. The results predicted by correct models are in bold lines.

Allen's model is also widely used to predict electrical transport coefficients [21,22,49,50]. The details of Allen's model can be found in Appendix C. The predicted electrical conductivities by Allen's model are shown as the blue dotted lines in FIG. 4. Allen's model matches MRTA model well in the temperature range of 100-300 K. However, it has significant deviation from MRTA as the temperature is below 60 K for Cu and 80 K for Al.

The electronic thermal conductivity of Cu and Al is shown in FIG. 5. It first decreases dramatically with temperature before 40 K, and then tends to a constant value as the temperature is high. Electronic thermal conductivity is related to electronic heat capacity and energy relaxation time. The heat capacity linearly increases with temperature. The energy relaxation time decreases dramatically with temperature in the low-temperature range, as it is shown in FIG. 5. Hence, the electronic thermal conductivity decreases quickly with temperature. At higher temperatures, the energy relaxation time is almost inversely proportional to the temperature due to the increase of scattering phase space induced by phonon distribution increasing, resulting in the smooth thermal conductivity curves. The electronic thermal conductivity predicted by ERTA matches the experimental data [6] well in the whole temperature range. For Cu, the difference between ERTA and MRTA is very small as the temperature is larger than 40 K. While the difference is significant in the whole temperature range for Al. This can be similarly interpreted as we have done in electrical conductivity.

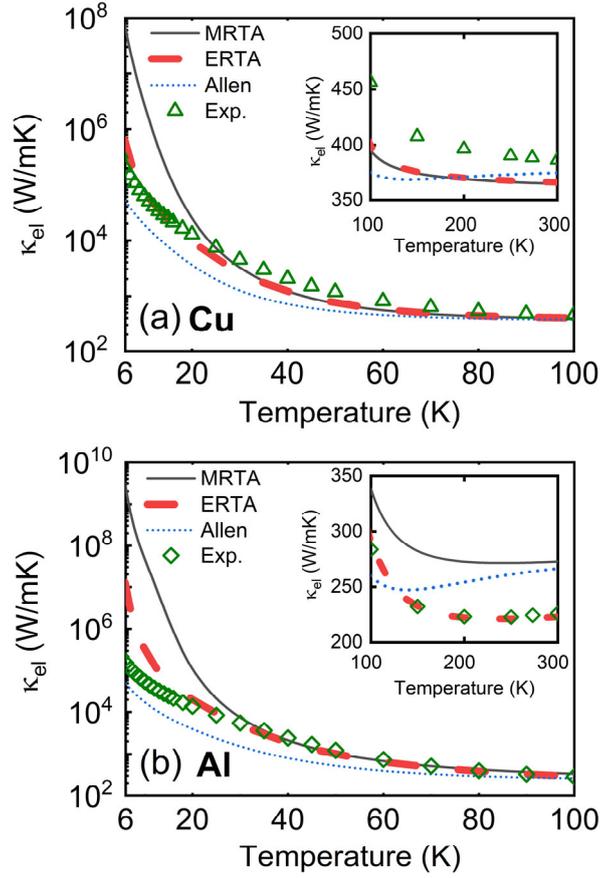

FIG. 5. The electronic thermal conductivity for (a) Cu and (b) Al calculated with MRTA (black solid lines), ERTA (red dash lines) and Allen's model (blue dot lines). The experimental data are taken from Ref. [6]. The insets of (a) and (b) are the electronic thermal conductivity in the temperature range of 100-300 K. The ideal experimental data is obtained by subtracting the residual resistance. The results predicted by correct models are in bold lines.

The electronic thermal conductivities predicted by Allen's model are also presented. Allen's model can capture the variation of electronic thermal conductivity for both Cu and Al. However, it has significant deviations from ERTA in the whole temperature range. Allen's model only employs the properties of electrons at the Fermi surface. Actually, the electron modes with energy go in the range of $\varepsilon_F \pm \mathrm{d}E$ contribute to the electronic thermal conductivity and the electron scattering rate cannot be approximated as a constant at the Fermi surface, as shown in FIG. 3. The variation of electrical and thermal conductivity at 100 K with a width of the Fermi window ($\mathrm{d}E$) is shown in FIG. A3. To ensure the convergence, the Fermi window width is set as 0.1 eV as the temperature is smaller than 100K, and it is set as 1.0 eV for the temperature range of 100-300 K.

## C. Electron relaxation times

To further analyze the differences in the results predicted by MRTA and ERTA, the averages of the two-electron relaxation times are present. They are given by [51]

$$\begin{cases} \langle \tau_\sigma \rangle = \dfrac{\sum\limits_{km} \dfrac{\partial f^0_{km}}{\partial \varepsilon_{km}} |v_{km}|^2 \tau_{\sigma,km}}{\sum\limits_{km} \dfrac{\partial f^0_{km}}{\partial \varepsilon_{km}} |v_{km}|^2} \\ \langle \tau_\kappa \rangle = \dfrac{\sum\limits_{km} \dfrac{\partial f^0_{km}}{\partial \varepsilon_{km}} |v_{km}|^2 \tau_{\kappa,km}}{\sum\limits_{km} \dfrac{\partial f^0_{km}}{\partial \varepsilon_{km}} |v_{km}|^2} \end{cases} \quad (12)$$

with $|v_{km}|^2 = (v^x_{km})^2 + (v^y_{km})^2 + (v^z_{km})^2$.

The momentum and energy relaxation times both show a similar temperature dependence for both Cu and Al, as shown in FIG. 6. They hold a rapid decline as the temperature is smaller than 40 K and then the decreasing trend slows down in the higher temperature range. The magnitude of momentum relaxation time is significantly larger than that of energy relaxation time for Cu at low temperatures. This indicates that the efficiency factor has significant effects in this temperature range. To quantify the effects of the efficiency factor, the average efficiency factor $\langle \alpha \rangle$ is given by

$$\langle \alpha \rangle = (1/\langle \tau_\sigma \rangle)/(1/\langle \tau_\kappa \rangle) \quad (13)$$

As can be seen in the inset of FIG. 6(a), the average efficiency factor increases with temperature below 40 K for Cu and converges to ~1.0 in the higher temperature range. While the difference between electrical and energy relaxation time of Al is smaller in low-temperature range. The efficiency factor is ~0.5 at 6 K for Al while ~0.1 for Cu. It should be noted that the efficiency factor is close to one in the vicinity of 40 K and there exists a slight fall as the temperature is higher in Al, which is different from the variation trend of Cu.

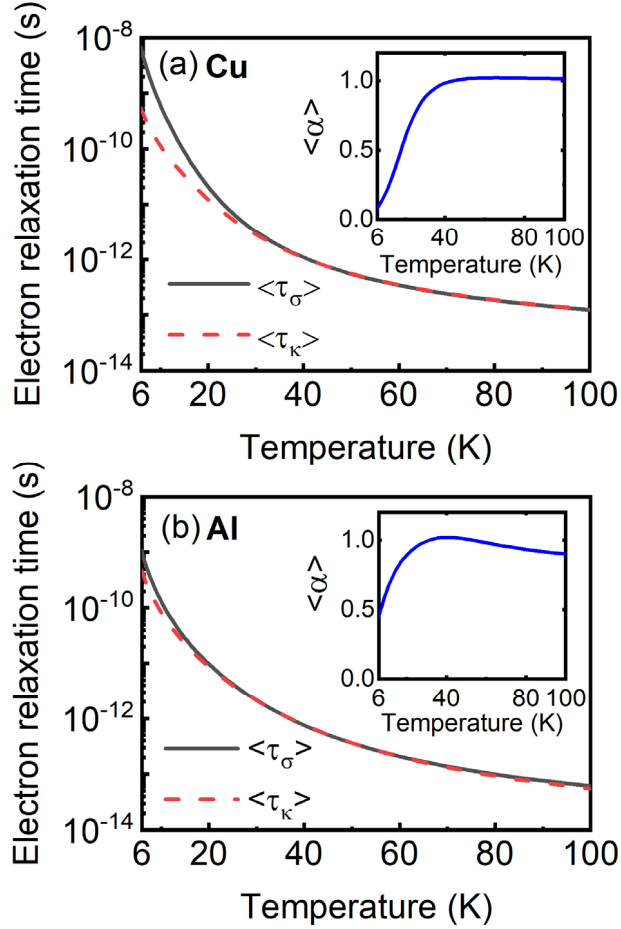

FIG. 6. The average momentum relaxation time $\langle \tau_\sigma \rangle$ and energy relaxation time $\langle \tau_\kappa \rangle$ with different temperatures for (a) Cu and (b) Al. The insets are the average efficiency factor $\langle \alpha \rangle$ for Cu and Al, respectively.

**D. Lorenz ratio**

The temperature-dependent Lorenz ratios with different models are presented in FIG. 7. According to the kinetic theory of electron [5], the Lorenz ratio for metals should be Sommerfeld value, as $L_0 = \pi^2 k_B^2 / (3e^2) = 2.44 \times 10^{-8} \, \text{W}\Omega/\text{K}^2$. This is based on the constant relaxation time assumption for free-electron metal. However, based on our calculation (momentum relaxation time for electrical conductivity and energy relaxation time for electronic thermal conductivity), the Lorenz ratio is no longer a constant at intermediate temperatures. It approaches zero at low temperatures and increases with temperature going toward $L_0$. Our predicted Lorenz number is consistent with the existing theory [4], which states that Lorenz number should be very small in low-temperature range and increases with temperature if only electron-phonon scattering considered. The Lorenz ratio predicted by constant relaxation time is almost unchanged with temperature and very close to the Sommerfeld value. The Lorenz ratios predicted by ERTA and MRTA also have large variations with temperature, which indicates the variation is induced by the mode-dependent electron relaxation time.

The Lorenz ratios predicted by ERTA and MRTA is incomparable with present work in the low-temperature range, which further confirms only momentum relaxation is used for electrical conductivity and energy relaxation time for thermal conductivity can capture the correct Lorenz ratio. However, these two models predict similar Lorenz ratios as present work when the temperature is larger than 40 K for Cu. This is because the two relaxation times do not have much difference in that temperature region. For Al, the Lorenz ratio predicted by ERTA and MRTA approaches that of TRTA only in a small range close to 40 K and large deviations exist in other temperature range. The momentum relaxation time and energy relaxation time are close to each other only in the vicinity of 40 K. Therefore, the further Lorenz ratio difference between present work and single relaxation time models is induced by the large difference in the two relaxation times $\tau_\sigma$ and $\tau_\kappa$. It should be noted the Lorenz ratios, either by MRTA or ERTA deviates from Sommerfeld value at low temperatures due to the electron mode dependence electron relaxation time, which has been explained in Ref. [4].

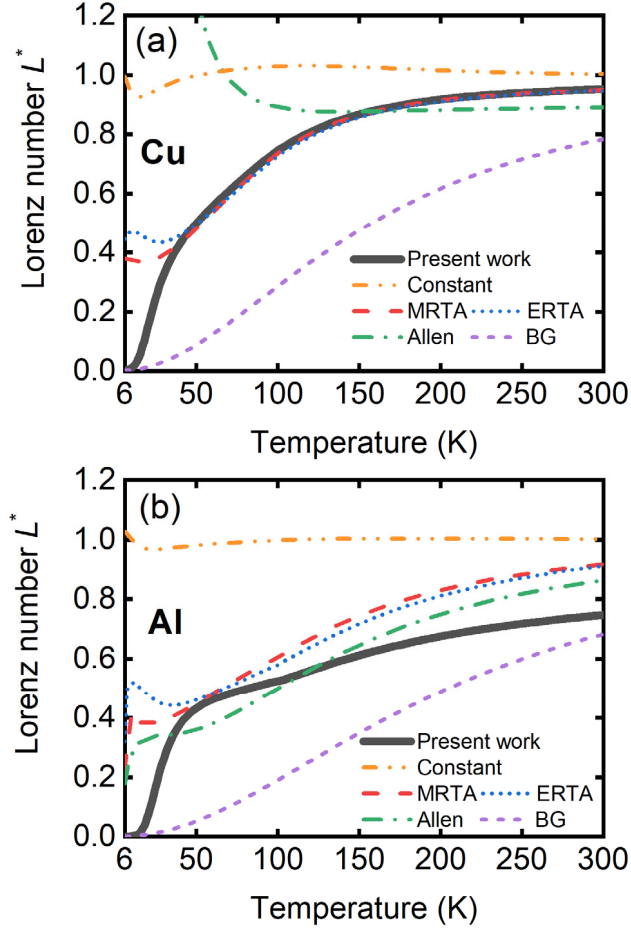

FIG. 7. The temperature dependent Lorenz ratio predicted in present work, by constant relaxation time (Constant), MRTA, ERTA, Allen's model (Allen) and BG model for (a) Cu and (b) Al. The Lorenz ratio is normalized by Sommerfeld value, as $L_0 = 2.44 \times 10^{-8} \, W\Omega/K^2$.

The Lorenz ratios predicted by Allen's model and BG model are also shown in FIG. 7. Allen's model cannot replicate the Lorenz ratio at low temperature range, while the departure is not that large as the temperature is high. The Lorenz ratio predicted by BG model holds the similar variation tendency with our present method, which is attributed to the two relaxation times are also employed in the BG model. Also, the Lorenz ratio predicted by the BG model converges to one as the temperature is high. However, it does not match our first-principles calculation in quantity. This is attributed to the strong assumptions in BG model.

### E. Inelastic electron-phonon scattering

The anomalous behavior of Wiedemann-Franz law at low temperatures in FIG. 7 can be further attributed to the inelastic electron-phonon scattering, which induces the difference between momentum relaxation time and energy relaxation time. The electron scattering processes for electrical current and heat current are different [5,52], as shown in FIG. 8. The electrons driven by the external electric field can only be scattered by large-angle scatterings (horizontal process) and return to equilibrium, which can change the direction of motion. In comparison, the electrons driven by external temperature gradient can be effectively scattered by both large-angle scatterings and small-angle scatterings (vertical process). As the large angle scatterings change the momentum of the electrons, it is also called quasi-elastic scattering. One should note there always exists energy transport in an electron-phonon scattering process. Strictly speaking, both large-angle scatterings and small-angle scatterings are inelastic.

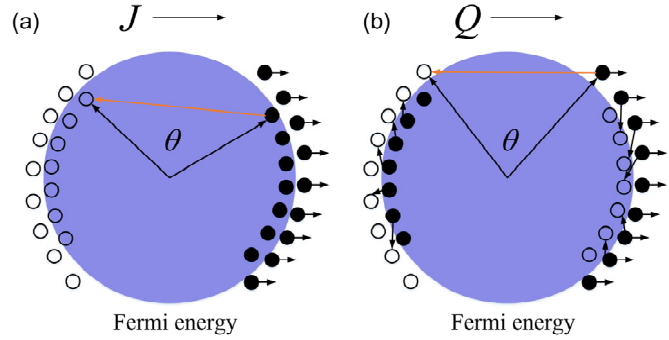

FIG. 8. Schematic representation of electron scattering in (a) an electric field and (b) under a temperature gradient. The electrons close to the Fermi surface are driven by the external perturbations. As for the external electrical field, the electrons can only return to equilibrium through large-angle scatterings (horizontal process), as illustrated by the orange arrow in (a). As for the external temperature gradient, the electrons can return to equilibrium through both large angle scatterings (orange arrow in (b)) and small-angle scatterings (vertical process), as illustrated by the small black arrows in (b). The filled small spheres are occupied electron states and the open small spheres are unoccupied electron states.

At low temperature, the electron scattering from long wavevector phonons are largely suppressed and the short wavevector phonons are the dominant scattering source for electrons. Short wavevector phonons cannot easily switch the direction of electron motion, but they do affect the energy transport of electrons by inelastic scattering. As a result, the momentum relaxation time is larger compared to energy relaxation time. To further clarify this mechanism, the mode-level efficiency factors $\alpha_{n\mathbf{k}}$, as $(1/\tau_{\sigma,n\mathbf{k}})/(1/\tau_{\kappa,n\mathbf{k}})$ are presented in FIG. 8. The efficiency factors at 10 K for the electron modes close to Fermi energy is significantly smaller than one for both Cu and Al, which indicates the significant inelastic scattering. There is almost no difference between 100 K and 50 K for Cu and the value approaches one, implying that the quasi-elastic scattering dominates. The efficiency factor of Al has some deviation from one at 50 K and 100 K. Note that the value at 100 K is slightly smaller compared to 50 K. This can interpret the decreasing trend of the average efficiency factor at higher temperatures in the inset of FIG. 6(b).

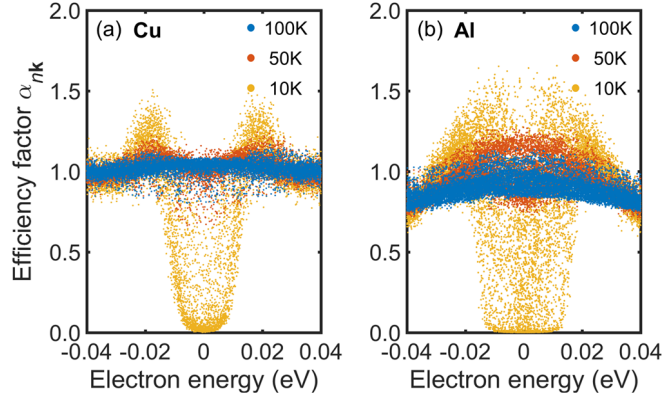

FIG. 9. The efficiency factor at electron mode $n\mathbf{k}$ with different temperatures for (a) Cu and (b) Al. The electron energy is normalized to Fermi energy ($\varepsilon_F$).

The Fermi surfaces with superimposed efficiency factor $\alpha_{n\mathbf{k}}$ is shown in FIG. 10. Cu holds quasi-spherical Fermi surface while Al owns an irregular Fermi surface. There is only one sheet for Cu and two sheets for Al. The Fermi surface of Cu is far from the Brillouin zone boundary while that of Al is very close to the boundary. These differences in Fermi surfaces cause the different electrical transport properties in Cu and Al. It is clear to see the anisotropy of the efficiency factor on the Fermi surfaces. The blue ribbons on Fermi surfaces at 10 K in FIG. 10 for Cu and Al are corresponding to the extremely small values in FIG. 9. The efficiency factor is almost uniform and approaches one over the whole Fermi surface at 50 K and 100 K for Cu. The electron modes far away from the Brillouin zone boundary on the inner sheet of the Femi surface for Al hold smaller efficiency factors. Most regions of the Fermi surface for Al hold smaller efficiency factors at 100 K compared to 50 K, which eventually results in the slightly smaller average efficiency factor at 100 K, as shown in the inset of FIG. 6(b).

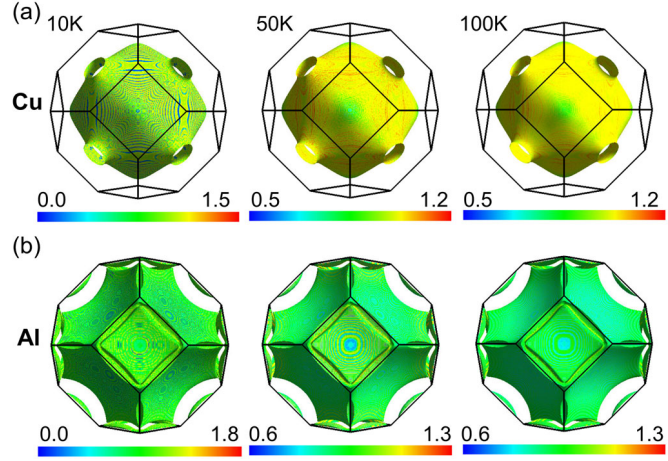

FIG.10. Fermi surfaces showing the efficiency factor at electron mode $n\mathbf{k}$ at 10 K, 50 K and 100 K for (a) Cu and (b) Al. There is only one sheet of Fermi surface for Cu, while two sheets for Al. The Fermi surfaces are plot with FermiSurfer package [53].

### F. Phonon thermal conductivity

The phonon thermal conductivity ($\kappa_{ph}$) for both Cu and Al at the temperature range of 6-300 K is shown in FIG. 11. Both phonon-phonon and phonon-electron scattering are included in the estimation of $\kappa_{ph}$. As we can see, the phonon thermal conductivity first increases with temperature and then decreases with temperature. The increase in low temperature is related to the increase of phonon heat capacity. Phonon-electron scattering is the main phonon scattering source in this region and it is weakly dependent on the temperature [54]. It would have a significant effect on the phonon thermal conductivity as the temperature is smaller than 100 K, as shown in Appendix E. The decrease in higher temperature is mainly attributed to the increase of anharmonic phonon-phonon scattering. Note $\kappa_{ph}$ occupy only a small proportion of the total thermal conductivity, as is smaller than 5% in the whole temperature range. The Lorenz ratio would have a rare change if the total thermal conductivity is employed in its estimation.

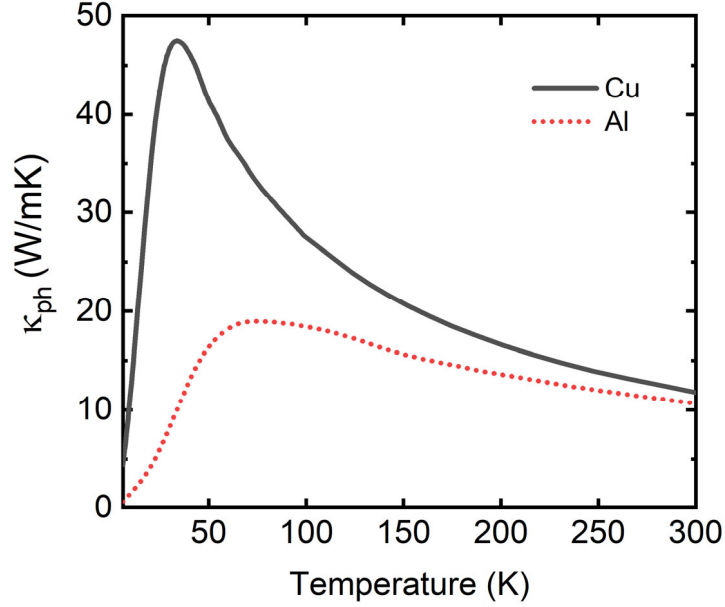

FIG. 11. The phonon thermal conductivity for Cu (solid line) and Al (dot line) in the temperature range of 6 to 300 K. Both phonon-phonon and phonon-electron scattering effects are included.

## IV. Conclusions

In summary, we perform a rigorous first-principles study on the thermal conductivity and Lorenz ratio for Cu and Al at intermediate temperatures. It is found small smearing broadening parameter, extremely dense **k**-point and **q**-point mesh should be used in order to obtain the correct electrical transport coefficients at intermediate temperatures. Importantly, it is shown that momentum relaxation time should be used for electrical conductivity and energy relaxation time for electronic thermal conductivity. There would exist large deviations in Cu as the temperature below 40 K and non-negligible deviations even at room temperature in Al if the relaxation times are misused. It is found there is an intrinsic deviation of the Lorenz ratio from the Sommerfeld value as the temperature is smaller than 40 K for Cu and almost the whole temperature range for Al, which is attributed to the considerable inelastic electron-phonon scattering. Finally, the phonon thermal conductivity in the pure metals is shown to be significantly smaller than the electron component. The presented calculation scheme can quantitatively obtain the phonon-limited thermal conductivity and Lorenz ratio, which is a relatively unexplored area. This approach can also be applied to other metallic systems and would enable the design of high performance metallic materials.

## Appendix

### A. Phonon dispersion and electron band structure

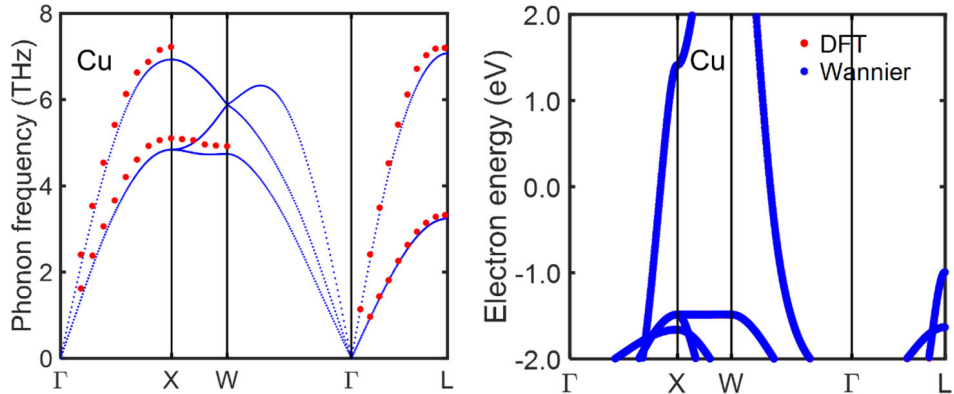

FIG. A1. (a) The phonon dispersion curve of Cu by DFT and experimental data. The experimental data (red dots) are taken from Ref. [55]. (b) The electron band structure. Red (DFT), blue (Wannier). The electron energy takes the Fermi energy ($\varepsilon_F$) as the reference.

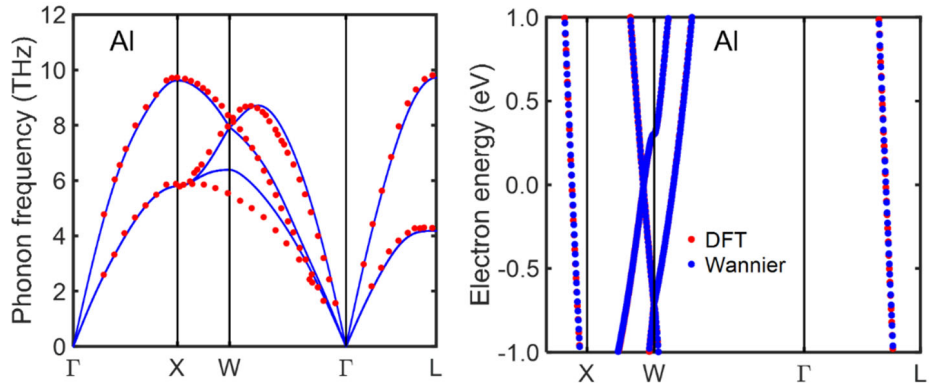

FIG. A2. (a) The phonon dispersion curve of Al by DFT and experimental data. The experimental data (red dots) are taken from Ref. [56]. (b) The electron band structure. Red (DFT), blue (Wannier). The electron energy takes the Fermi energy ($\varepsilon_F$) as the reference.

## B. Convergence test for Fermi window width

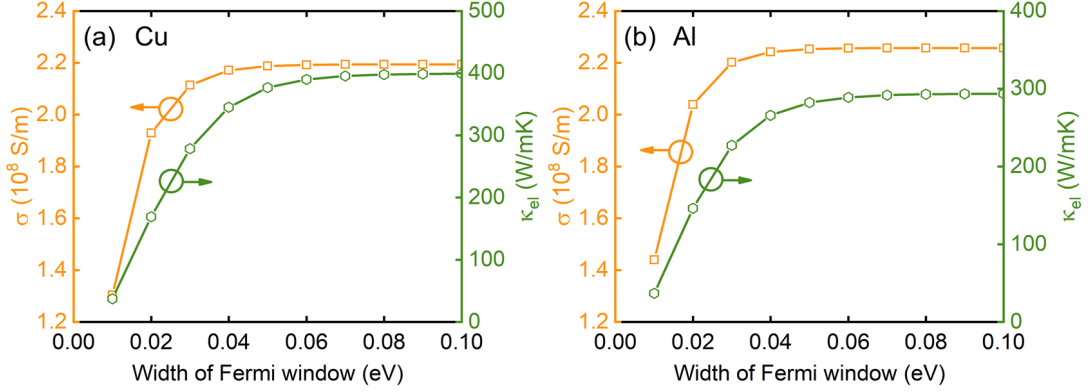

FIG. A3. Calculated electrical and thermal conductivity for (a) Cu and (b) Al at 100 K with respect to the width of the Fermi window.

## C. Allen's model for transport coefficients

The electrical resistivity is given by [18,19]

$$\rho_{el}(T) = \frac{1}{\sigma(T)} = \frac{2\pi V k_B T}{e^2 \hbar N_F \langle v_\alpha^2 \rangle} \int_0^\infty \frac{d\omega}{\omega} \frac{x^2}{\sinh^2 x} \alpha^2 F_{tr}(\omega) \qquad (A1)$$

$N_F$ is the electron density of state per spin and per unit cell at the Fermi surface. $\alpha$ is the coordinate of cartesian system. $\langle v_\alpha^2 \rangle$ means the average velocity square in coordinate $\alpha$. $x = \hbar\omega / 2k_B T$ is a dimensionless parameter. The Eliashberg transport function $\alpha^2 F_{tr}$, which is expressed as [32]

$$\alpha^2 F_{tr}(\omega) = \frac{1}{N(\varepsilon_F)} \sum_{qj} \sum_{knm} |g^{qj}_{k+qm,kn}|^2 \delta(\hbar\omega - \hbar\omega_{qj}) \times \delta(\varepsilon_{kn} - \varepsilon_F)\delta(\varepsilon_{k+qm} - \varepsilon_F) \alpha_{k+qm,kn} \qquad (A2)$$

where $\alpha_{k+qm,kn}$ is the efficiency factor in Eq. (6).

The thermal resistivity is given by [18,19]

$$\rho_{th}(T) = \frac{1}{\kappa_{el}(T)} = \frac{1}{L_0 T} \frac{2\pi V k_B T}{e^2 \hbar N_F \langle v_\alpha^2 \rangle} \int_0^\infty \frac{d\omega}{\omega} \frac{x^2}{\sinh^2 x} \left\{ \left[1 - \frac{2}{\pi^2} x^2 \right] \alpha^2 F_{tr}(\omega) + \frac{6}{\pi^2} x^2 \alpha^2 F(\omega) \right\} \quad (A3)$$

where $L_0$ is the Sommerfeld value of Lorenz number. $\alpha^2 F$ is the Eliashberg spectral function, which is read as[32]

$$\alpha^2 F(\omega) = \frac{1}{N(\varepsilon_F)} \sum_{\mathbf{q}j} \sum_{\mathbf{k}nm} \left| g_{\mathbf{k}+\mathbf{q}m,\mathbf{k}n}^{\mathbf{q}j} \right|^2 \delta(\hbar\omega - \hbar\omega_{\mathbf{q}j}) \times \delta(\varepsilon_{\mathbf{k}n} - \varepsilon_F) \delta(\varepsilon_{\mathbf{k}+\mathbf{q}m} - \varepsilon_F) \quad (A4)$$

**D. Lorenz ratio by BG model**

The Bloch-Grüneisen model gives the Lorenz number as

$$L = \frac{L_0}{1 + \frac{3}{\pi^2} \left(\frac{k_F}{q_D}\right)^2 \left(\frac{\theta_D}{T}\right)^2 - \frac{1}{2\pi^2} \frac{J_7(\theta/T)}{J_5(\theta/T)}} \quad (A5)$$

where $k_F$ and $q_D$ are Fermi wave vector and Debye wave vector, respectively. $\theta_D$ is Debye temperature, which is predicted as 322 K for Cu and 446 K for Al from first principles. $J_n$ ($n$ is an integer) is defined as

$$J_n\left(\frac{\theta}{T}\right) \equiv \int_0^{\theta/T} \frac{x^n e^x}{(e^x - 1)^2} dx \quad (A6)$$

**E. Phonon-electron scattering effects on phonon thermal conductivity**

The phonon thermal conductivity is given by

$$\kappa_{\text{ph},\alpha\beta} = \sum_{\lambda} c_{v,\lambda} v_{\lambda,\alpha} v_{\lambda,\beta} \tau_{\lambda} \qquad (A7)$$

where $\lambda$ is the phonon mode denotation. $c_v$ is the phonon heat capacity, $v$ is the phonon group velocity, and $\tau$ is the phonon relaxation time. All the three parameters are phonon mode dependent. According to Matthiessen's rule, the phonon relaxation time can be expressed as $1/\tau_{\lambda} = 1/\tau_{\lambda}^{PPI} + 1/\tau_{\lambda}^{PEI}$. Here, $1/\tau_{\lambda}^{PPI}$ and $1/\tau_{\lambda}^{PEI}$ the phonon scattering induced by phonon-phonon and phonon-electron scattering.

To demonstrate the effects of phonon-electron scattering, the phonon thermal conductivity without and with phonon-electron scattering for Cu is shown in FIG. A4.

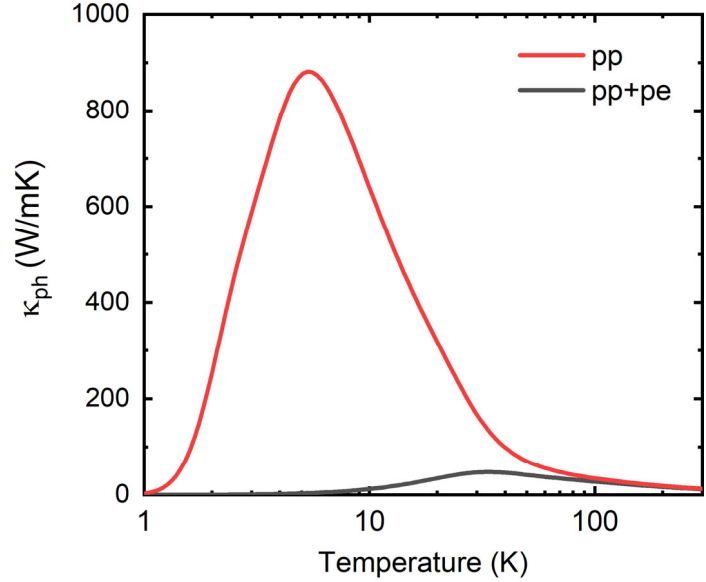

FIG. A4 The phonon thermal conductivity without (pp) and with (pp+pe) phonon-electron scattering for Cu.

## Acknowledgments

We would like to thank Dr. Xiaokun Gu and Dr. Wu Li for valuable discussions. This work was supported by the National Natural Science Foundation of China No. 51676121 (HB). Simulations were performed with computing resources granted by HPC (π) from Shanghai Jiao Tong University.